\documentstyle[aps,epsf]{revtex}  

\begin{document}        

\baselineskip 14pt
\title{New DELPHI Results from Semileptonic $b,~c$ Decays}
\author{M. Margoni}
\address{Dipartimento di Fisica, Universit\`a di Padova and INFN,
Via Marzolo 8, I-35131 Padua, Italy}
%
\maketitle              

\begin{abstract}        
The inclusive direct and cascade semileptonic branching ratios of the $b$ quark
were measured from a fit to the lepton momentum spectrum in the $B$-hadron rest frame, exploiting 
the charge correlation between the lepton and the heavy meson:
\begin{eqnarray*}
BR(\bar{b}\rightarrow l^+)&=&10.32\pm0.12(stat.)\pm0.27(syst.)^{+0.31}_{-0.21}(model)\%\\ 
BR(\bar{b}\rightarrow \bar{c}\rightarrow l^-)&=&7.32\pm0.26(stat.)\pm0.32(syst.)^{+0.06}_{-0.16}(model)\%\\ 
BR(\bar{b}\rightarrow c\rightarrow l^+) &=&0.94\pm0.16(stat.)\pm0.28(syst.)^{+0.33}_{-0.50}(model)\%
\end{eqnarray*}
The value of the $|V_{cb}|$ element in the CKM quark mixing matrix was computed 
from the differential decay width of the process $B^0\rightarrow D^{*+}l^-\nu$, measured as a function
of the product $\omega$ of the $B$ and $D^*$ 4-velocities, using an inclusive method of $D^*$
reconstruction:
\begin{center}
$|V_{cb}|=41.70\pm2.35(exp.)\pm1.38(th.)\times 10^{-3}$
\end{center} 
A measurement of the ratio $\frac{|V_{ub}|}{|V_{cb}|}$ has been obtained by using the reconstructed
secondary hadronic mass in semileptonic decays of $B$ hadrons, and found to be:
\begin{center}
$\frac{|V_{ub}|}{|V_{cb}|}=0.104\pm0.012(stat.)\pm0.015(syst.)\pm0.009(model)$
\end{center}
The charm semileptonic branching ratio
was measured using a double tag method based on the detection of exclusively reconstructed $D$ mesons
accompanied by a lepton in the opposite hemisphere, giving:
\begin{center}
$BR(c\rightarrow l)=(9.58\pm0.42(stat.)\pm0.28(syst.))\%$
\end{center}
\
\end{abstract}   	

\section{Semileptonic Branching Ratios in $B$ decays}

Leptons in $b$-hadron decays can be produced both from direct $\bar{b}\rightarrow l^+$ and cascade 
$\bar{b}\rightarrow \bar{c} \rightarrow l^-$ decays which can be separated exploiting 
the correlation between the lepton charge and the $b$-hadron flavour.
The wrong sign charm decay $\bar{b}\rightarrow c\rightarrow l^+$, 
can be isolated from the direct decay 
by the low momentum signature of the leptons; the present analysis gives the first direct measurement of
$BR(\bar{b}\rightarrow c\rightarrow l^+)$.

From about $1.4$ million hadronic $Z^0$ decays collected during 1994, a $b\bar{b}$ event sample was selected
with efficiency $\epsilon^b\sim 50\%$ and purity $Pur^b\sim 95\%$
by means of the impact parameter, energy and angular distributions of the reconstructed charged tracks.

An inclusive $B$ reconstruction was performed from the high rapidity particles in the jet and a secondary vertex was fitted;
the energy and angular resolutions were $\sigma(E_B)\sim 7\%$ and $\sigma(\theta_B)\sim0.8^o$ respectively. 
The charge of the $B$ hadron was determined as
$Q_B=\sum Q^i\cdot Pr^i_b$
where $Q^i$ is the charge of the $i^{th}$ track and $Pr^i_b$ is the probability for the track to come from $B$ decay, 
computed using a neural network in terms of kinematical and topological variables. 

A second neural network was used in order to determine the $b$-hadron flavour, exploiting the charge of identified $K$ and
slow pions from $D^*$ decay, the jet charge and $Q_B$. 
A flavour discrimination $Pur(b\Leftrightarrow \bar{b})\sim85\%$ was obtained with an efficiency 
$\epsilon(b\Leftrightarrow \bar{b})\sim45\%$.

\begin{figure}[htb]
\vskip -.2 cm
\centerline{\epsfxsize 3.5truein \epsfysize 3.5truein \epsfbox{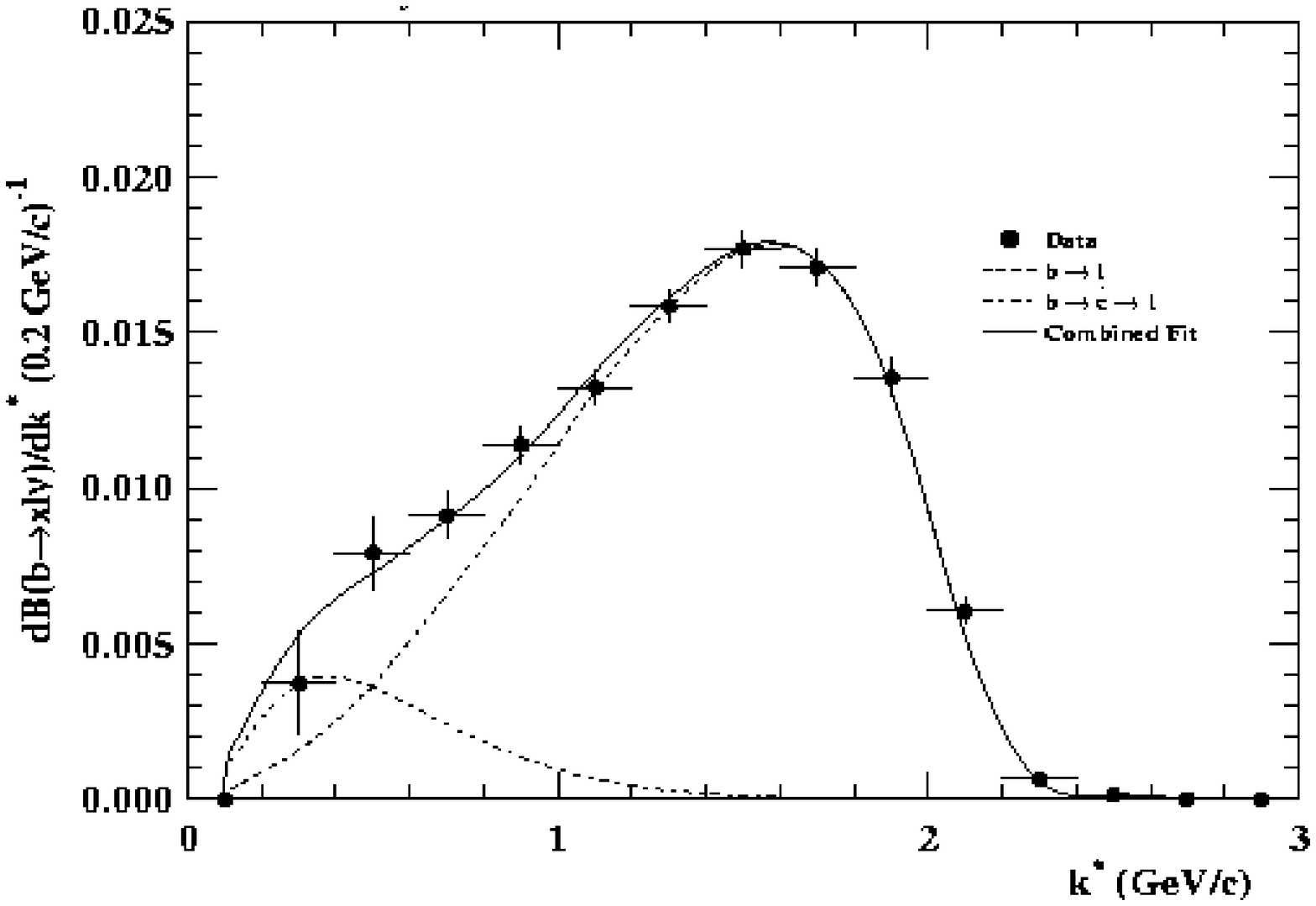}\epsfxsize 3.5truein \epsfysize 3.5truein \epsfbox{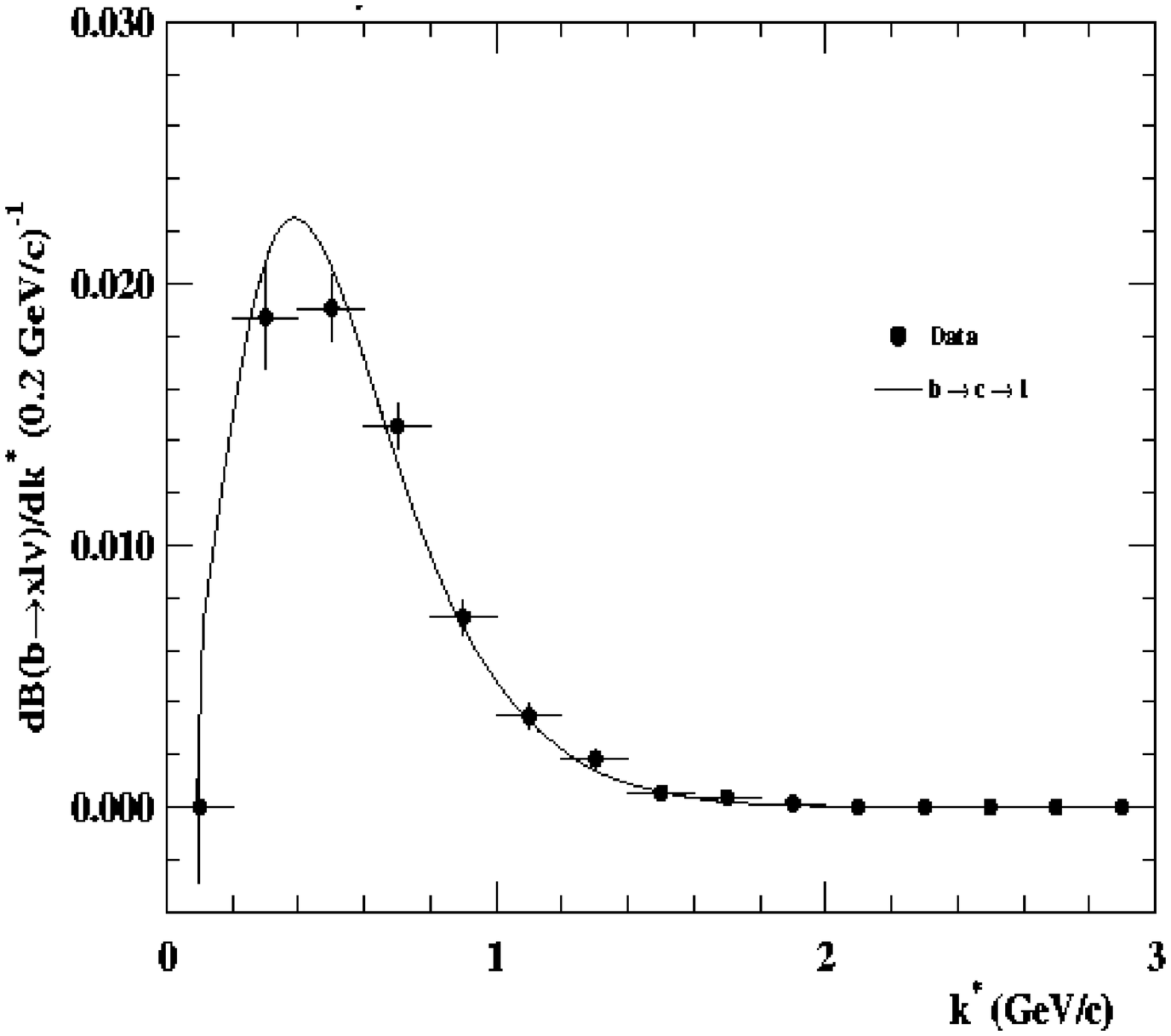}}   
\caption[]{
\label{fig:lfit}
\small Fitted $k^*_l$ spectrum in $B$ semileptonic decays. Left: same sign $(B,l)$, Right: opposite sign $(B,l)$.} 
\end{figure}

The lepton energy in the $b$-hadron rest frame $k^*_l$ was computed from the $B$ and the lepton 4-momenta with a resolution
$\sigma(k^*_l)\sim0.1~GeV/c$.
The background due to non $b$-events, fake $b$-hadrons and fake leptons was determined from the simulation and subtracted 
from the experimental $k^*_l$ distribution.
From a fit to $\frac{dN}{dk^*_l}$ the following preliminary results were obtained:
\begin{eqnarray}
\nonumber BR(\bar{b}\rightarrow l^+)&=&10.32\pm0.12(stat.)\pm0.27(syst.)^{+0.31}_{-0.21}(model)\%\\ \nonumber
\nonumber BR(\bar{b}\rightarrow \bar{c}\rightarrow l^-)&=&7.32\pm0.26(stat.)\pm0.32(syst.)^{+0.06}_{-0.16}(model)\%\\ \nonumber
\nonumber BR(\bar{b}\rightarrow c\rightarrow l^+) &=&0.94\pm0.16(stat.)\pm0.28(syst.)^{+0.33}_{-0.50}(model)\% \nonumber
\end{eqnarray}
The dominant systematics came from the lepton identification, the neural networks performances and the heavy flavour semileptonic
decay models.
Figure~\ref{fig:lfit} shows the results of the fit for the same and opposite sign $(B,~l)$ charge correlation samples
obtained using the ACCM model.

\section{$|V_{cb}|$ Measurement}

The Heavy Quark Effective Theory~\cite{iw} allows a precise determination of 
$|V_{cb}|$, the element of the CKM mixing matrix relating the 
$b$ and the $c$ quarks, from the analysis of the semileptonic decay 
$\bar{B^{0}}~\rightarrow D^{*+} l^{-} \bar{\nu}$. 
The following relation holds: 
\begin{equation}
\frac{d\Gamma}{d\omega} = \frac{1}{\tau_{B}}\cdot \frac{dBr}{d\omega} \propto G(\omega)\times A_1^2(\omega)\times |V_{cb}|^2
\label{eq:diff}
\end{equation}
where $\omega=v_{B}\cdot v_{D^*}$ is the product of the two heavy mesons
4-velocities and is related to the momentum transfer:
$q^2=(P_{\bar{B^0}}-P_{D^{*+}})^2=M^2_{B^0}+M^2_{D^*}-2M_{B^0}M_{D^*}\omega$.\par
$G(\omega)$ is a phase space function and $A_1(\omega)$ is a model dependent 
form factor, which can be expressed in terms of a single unknown parameter $\rho^2_{A_1}$~\cite{cln}:
\begin{equation} 
A_1(\omega) = A_1(1)\times {\cal F} (\rho^2_{A_1})
\label{eq:ff}
\end{equation}
In the limit in which the $D^*$ meson is
produced at rest in the $B$ reference frame 
$(\omega=1\leftrightarrow q^2=~10.7~GeV^2/c^4)$, the theory predicts\cite{s}
\begin{equation}
A_1(1)\approx 0.91\pm 0.03
\label{eq:ffval}
\end{equation} 
Experimentally, the goal is to measure $\frac{d\Gamma}{d\omega}$
at the point of zero recoil $\omega=1$.  
Due to the vanishing of phase space, the product
$A_1(1)\cdot |V_{cb}|$ was determined by an extrapolation from the 
$\frac{dN}{d\omega}$ distribution.

\begin{figure}[htb]
\centerline{\epsfxsize 3.5truein \epsfysize 4.truein \epsfbox{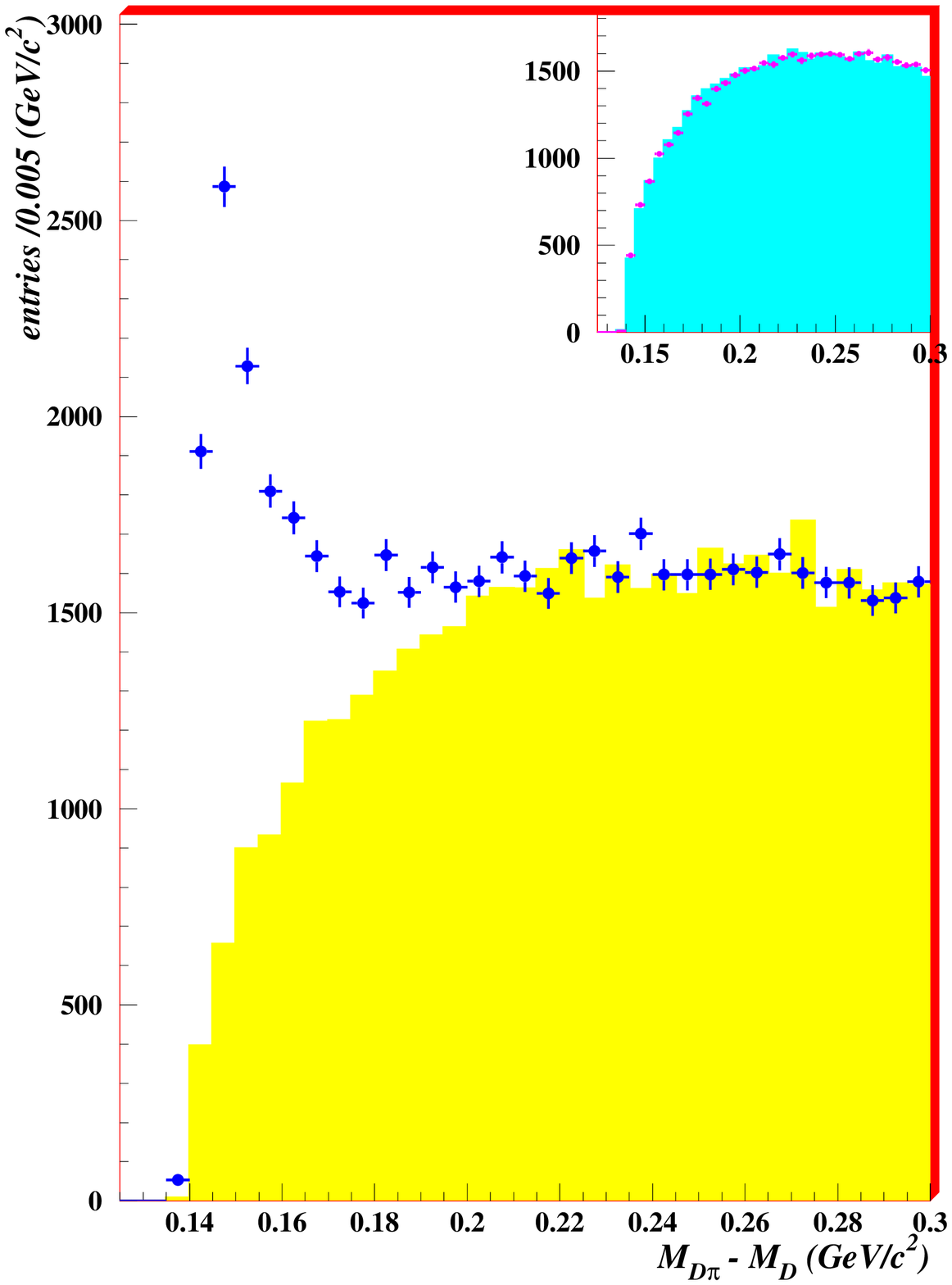}\epsfxsize 3.5truein \epsfysize 4.truein \epsfbox{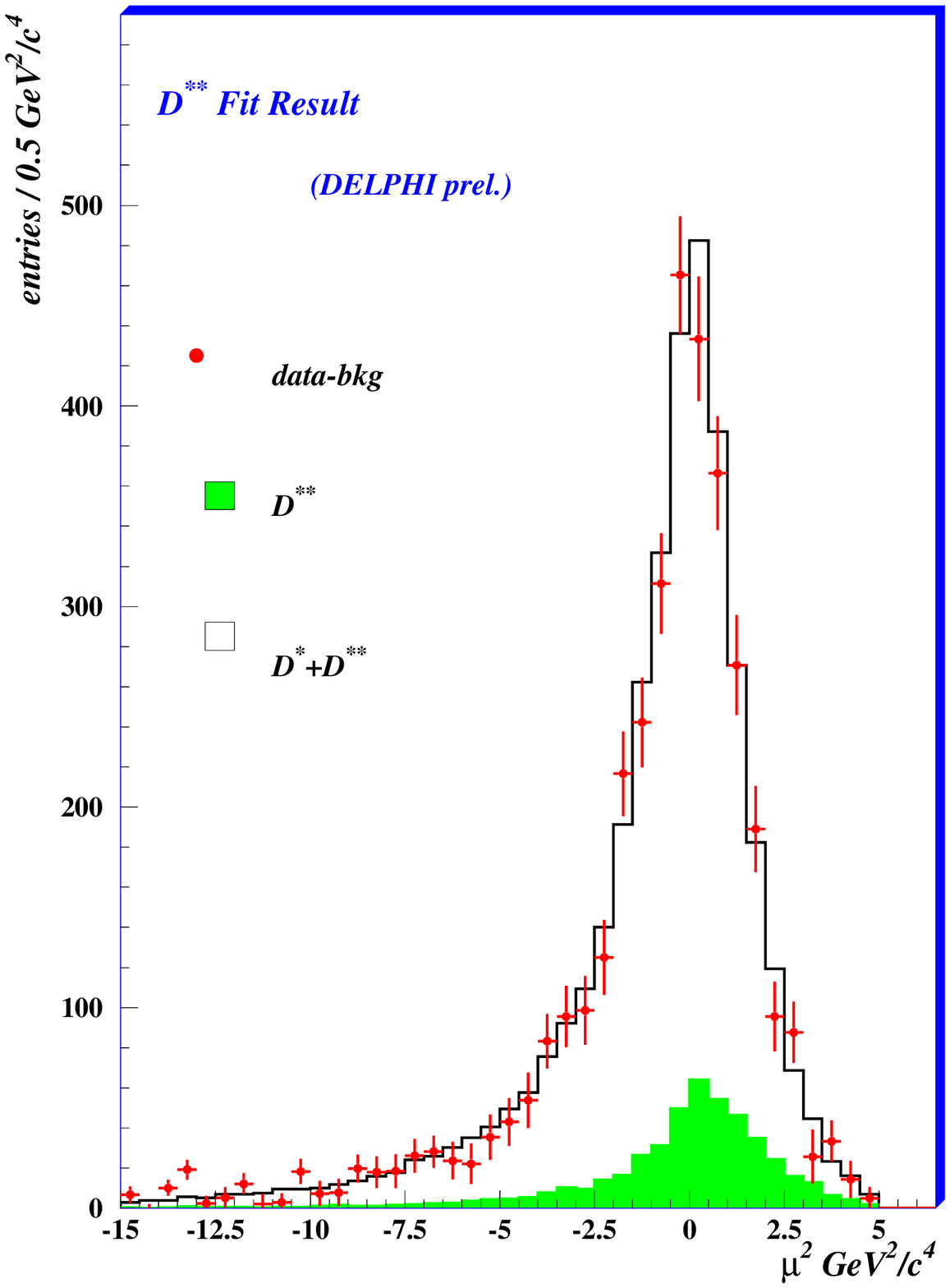}}   
\vskip -.2 cm
\caption[]{
\label{fig:dm1}
\small Left: Mass difference $M_{D^0\pi}-M_{D^0}$. Right charge (crosses), wrong charge (shaded area), 
normalised in the side band. Box: simulation: combinatorial from right charge (crosses), compared to wrong charge
sample; Right: fit to the $\mu^2$ distribution for the determination of the $D^{**}$ contamination.}
\end{figure}

The data collected from 1992 to 1995 were used in the analysis.
$D^0$ mesons were inclusively reconstructed from all the particles in the lepton jet, by means of their rapidity and
impact parameter. Pions with charge opposite to the lepton were then added to build a $D^*$, and events finally tagged
based on the mass difference $M_{D^0\pi}-M_{D^0}<0.165~GeV/c^2$. The combinatorial background shape was estimated from
the $(D^*,l)$ sample with the wrong charge correlation. A signal sample of
$N_{D^*l}=7075\pm164$ events was found with a selection efficiency 
$\epsilon=(14.7\pm0.1)\%$. Figure~\ref{fig:dm1} shows the $\delta m$ 
distribution for the final sample.
The 4-momentum of the $B$ was obtained from energy and momentum conservation and from the secondary
vertex reconstruction; the energy and angular resolutions were $\sigma(E_{B^0})\sim 10\%$ and 
$\sigma(\theta_{B^0})\sim\sigma(\phi_{B^0})\sim1^o$ respectively.
The dominant physics background was due to the semileptonic decays 
$B\rightarrow D^{**} l \nu,~D^{**}\rightarrow D^{*} X$. 
Signal events can be distinguished from the cascade ones
exploiting the Squared Recoil Mass $\mu^2=(P_{B^0}-(P_{D^*}+P_l))^2\equiv M^2_\nu$ which
should be zero for signal events, whereas
is usually greater than zero in case of background processes due to additional
particles produced in the $D^{**}$ decay. 
From a fit to the $\mu^2$ distribution with the $\mu^2(D^{**})$ shape fixed to HQET prediction\cite{mor},
the ratio\par
\begin{center}
$R^{**}=\frac{BR(b\rightarrow D^{**}l\nu)}{BR(b\rightarrow D^{**}l\nu)+BR(\bar{B^0}\rightarrow D^{*+}l\nu)}=0.18\pm0.11(stat.)\pm0.05(syst.)$\par
\end{center}
was determined. Figure~\ref{fig:dm1} shows the result of the fit.

\begin{figure}[htb]
\centerline{\epsfxsize 3.5truein \epsfysize 3.8truein \epsfbox{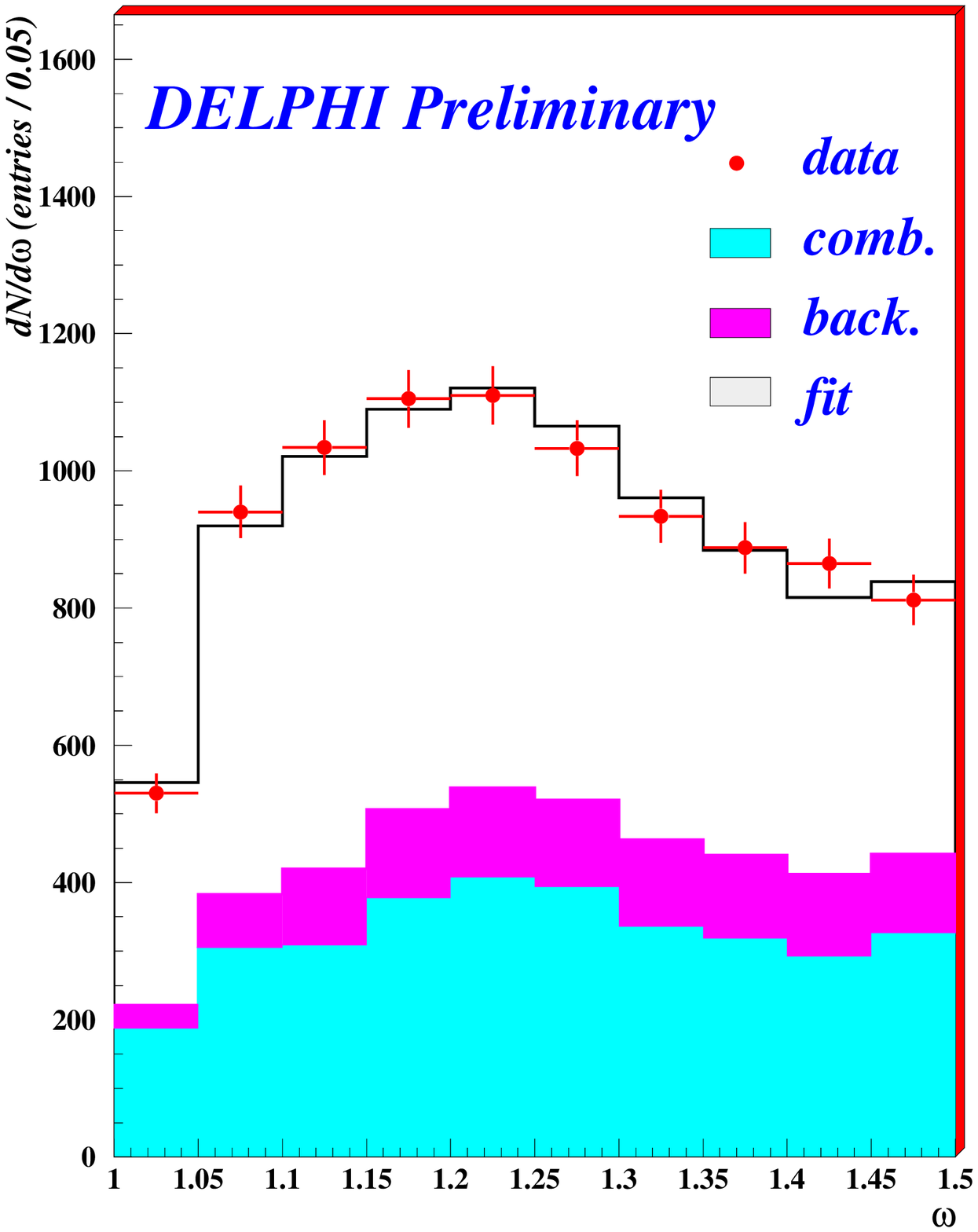}\epsfxsize 3.5truein \epsfysize 3.8truein \epsfbox{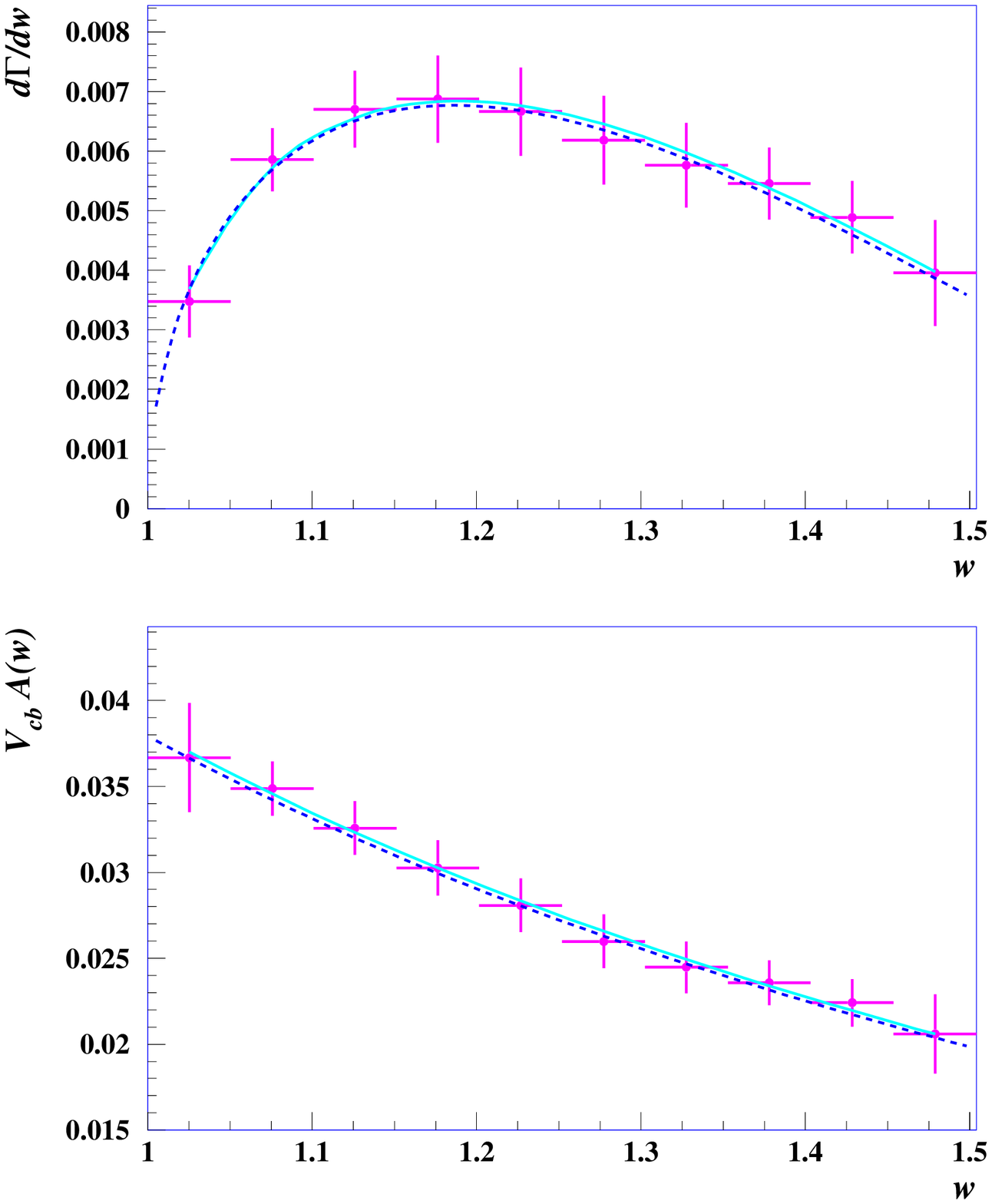}}   
\vskip -.2 cm
\caption[]{
\label{fig:vcb}
\small Left: fit to the $\omega$ distribution;
        Upper Right: unfolded differential decay width. Lower Right: unfolded decay form factor.}
\end{figure}

From the comparison of the $\omega$ distribution with the model function~\ref{eq:diff}, taking into account the 
experimental resolution and using the result~\ref{eq:ffval}, the values:
\begin{center}
$|V_{cb}| = (41.70\pm2.35(exp.)\pm1.38(th.))\times 10^{-3}$\par
$\rho^2_{A_1} = 1.39\pm0.13(stat.)\pm0.18(syst.)$
\end{center}
were obtained, where the dominant systematics come from  
$BR(b\rightarrow \bar{B^0}),~\tau_{\bar{B^0}},~BR(D^*\rightarrow D^0\pi)$, 
the experimental resolution and the $D^{**}$ fraction and decay models. 
Figure~\ref{fig:vcb} show the result of the fit together with the $\frac{d\Gamma}{d\omega}$ 
and the decay form factor unfolded distributions.
 
\section{$\frac{|V_{ub}|}{|V_{cb}|}$ from the Hadronic Recoil Mass Spectrum in $B$ semileptonic decays}

The measurement of the branching ratio for the decay $b\rightarrow u l \nu$ provides the most precise way
to determine the $|V_{ub}|$ element of the CKM mixing matrix. Evidence of the non-zero value of $|V_{ub}|$
has been obtained both by observing leptons produced in $B$ decays with momentum above the charm kinematic
threshold~\cite{endp} and by the measurement of the charmless exclusive decays 
$B\rightarrow \pi l \nu,~\rho l \nu$~\cite{excl}. 
These analyses were dominated by the theoretical error due respectively to the model dependence in the extrapolation
to low lepton momentum and to the uncertainty on the hadronic matrix elements.

According to HQET predictions at $O(1/m^3_b)$~\cite{ura}, $|V_{ub}|$ can be extracted from the inclusive width 
$\Gamma(B\rightarrow X_u l \nu)$ with a theoretical error $\delta(V_{ub})_{th}\sim 4\%$, dominated by the uncertainties on
the mass of the $b$ quark and $\alpha_s$.
Experimentally, a way to discriminate between the signal events and the dominant background due to the decay
$B\rightarrow X_c l \nu$ is provided by the Hadronic Recoil Mass $M_X$ which extends to higher values for the charmed final
states with respect to the charmless ones~\cite{barg}. According to recent calculations~\cite{bigi}, the signal region
can be defined by the cut $M_X<1.6~GeV/c^2$ which represents a good compromise between a reasonable high $b\rightarrow c l \nu$ 
rejection and a little model dependence in the calculation of the efficiency for the signal events 
$(\delta\epsilon(b\rightarrow u l \nu)\sim 10\%)$.

From the data collected between 1993 and 1995, a $b\bar{b}$ enriched sample was selected with purity $Pur^b\sim 65\%$ and
efficiency $\epsilon^b\sim 85\%$. The Hadronic Secondary System was then reconstructed with an efficiency of about $75\%$
from all the tracks classified as coming from $B$ decays by means of kinematical and topological variables.
The background from $b\rightarrow c l \nu$ was depleted according to the lepton impact parameter w.r.t
the secondary vertex, $K$ identification and 
$D^*$ inclusive reconstruction. The lepton energy in the $B$ rest frame was computed with a resolution 
$\sigma(E^*_l)\sim 14\%$. 

\begin{figure}[htb]
\centerline{\epsfxsize 3.5truein \epsfysize 3.truein \epsfbox{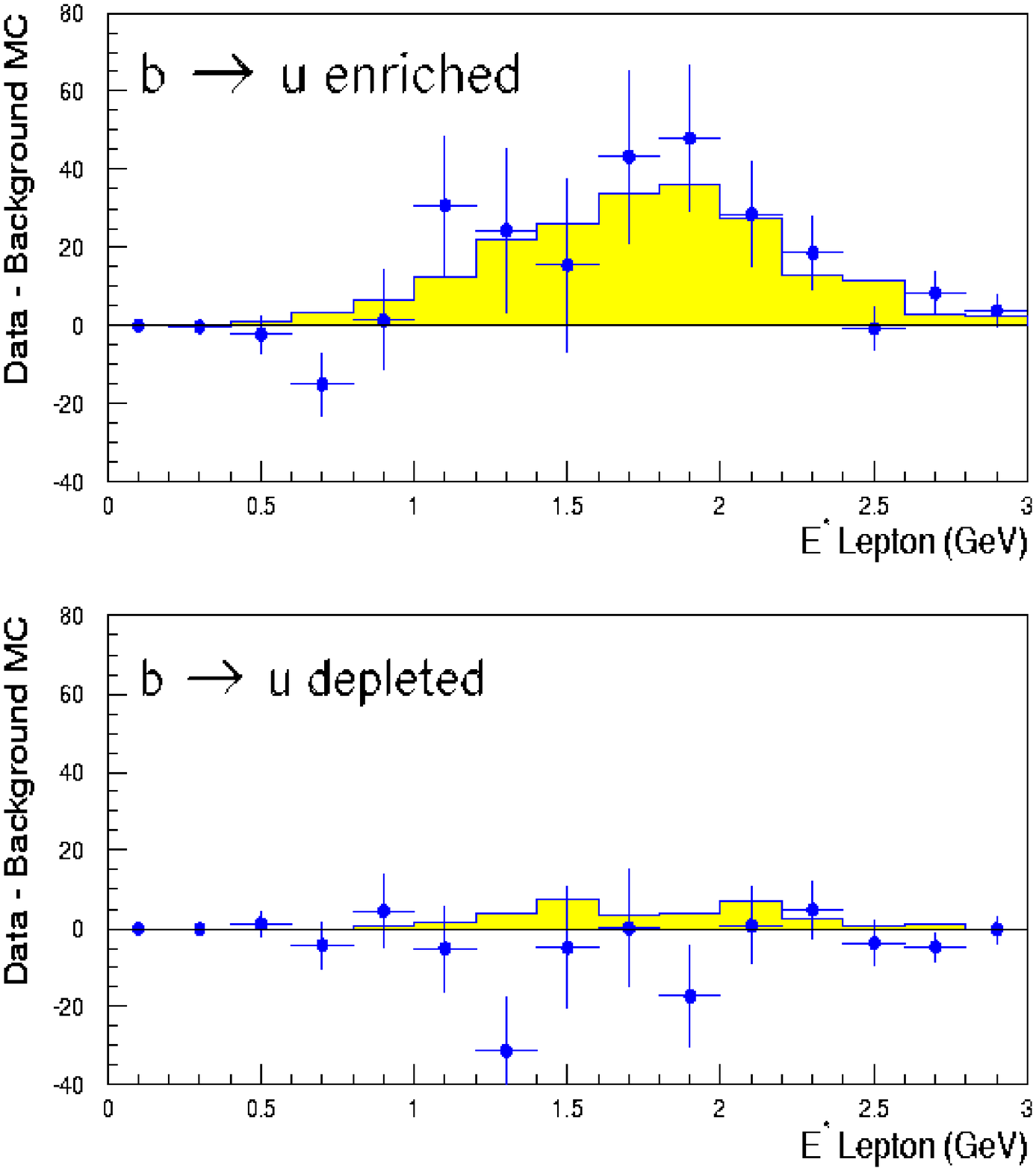}\epsfxsize 3.5truein \epsfysize 3.truein \epsfbox{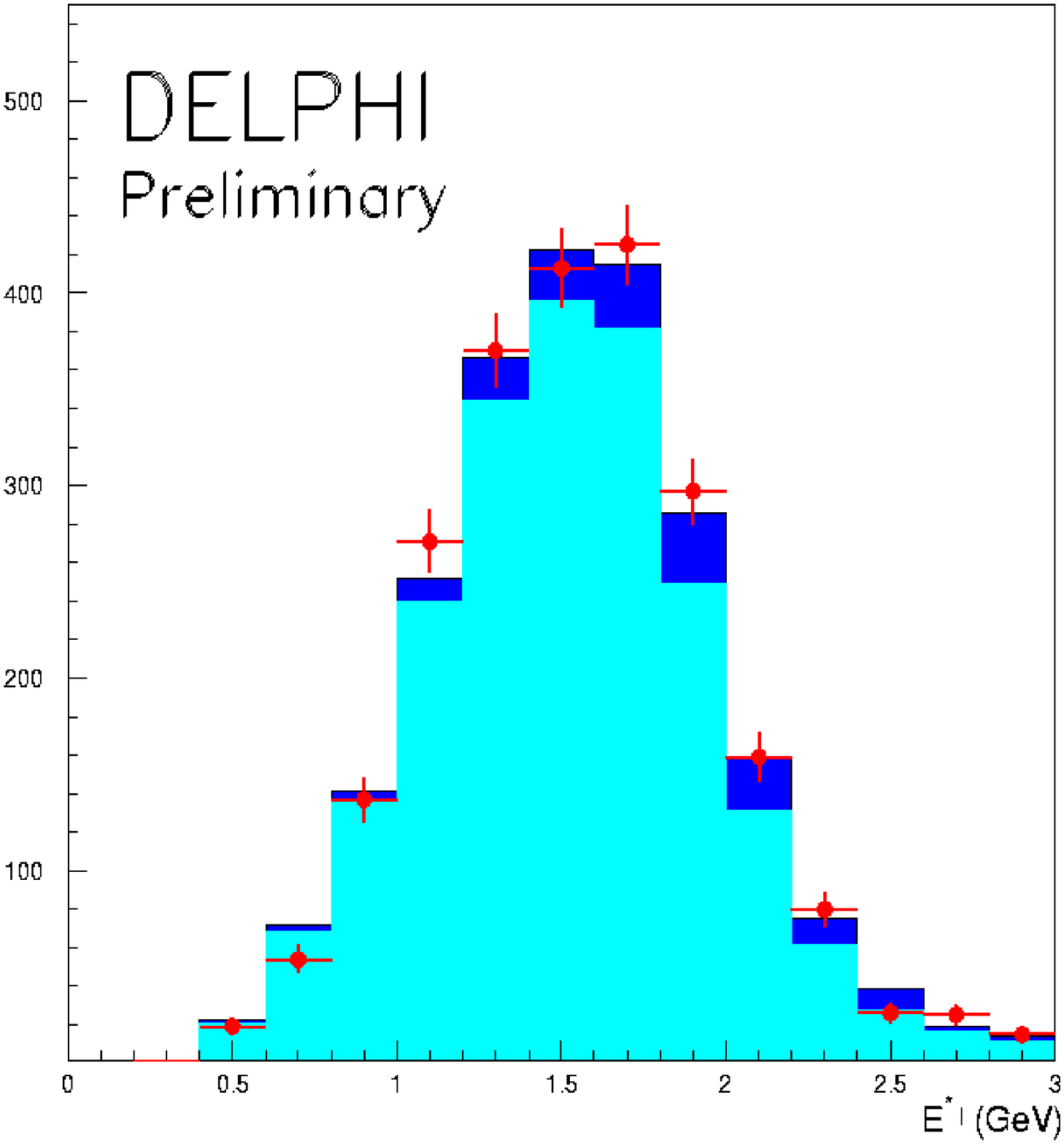}}   
\vskip -.2 cm
\caption[]{
\label{fig:vub}
\small Left: background subtracted $E^*_l$ distributions ($M_X<1.6~GeV/c^2$) for the enriched (upper plot) and 
depleted (lower plot) $b\rightarrow u l \nu$ samples; Right: fit to the $E^*_l$ distribution; the 
dark area shows the $b\rightarrow u l \nu$ fitted sample.}
\end{figure}

In the $b\rightarrow u$ enriched sample, an excess of $N_{b\rightarrow u l \nu}=205\pm 56$ events
was found in the data compared to the expected background; the selection efficiency for signal events was
$\epsilon\sim 11\%$.
Figure~\ref{fig:vub} shows the $E^*_l$
distribution after the background subtraction for the $b\rightarrow u$ enriched and depleted samples.
The fraction of $b\rightarrow u l \nu$ events in the selected sample is proportional to the ratio $\frac{|V_{ub}|}{|V_{cb}|}$.
From a fit to the $E^*_l$ distribution the preliminary value:
\begin{center}
$\frac{|V_{ub}|}{|V_{cb}|}=0.104\pm0.012(stat.)\pm0.015(syst.)\pm0.009(model)$ 
\end{center}
was extracted, where the systematics come from
the charm decay branching ratios, the $b$-hadron production and decay properties, the lepton and 
secondary system reconstruction efficiencies and the $b\rightarrow u l \nu$ modelisation. Figure~\ref{fig:vub} shows
the result of the fit.

\section{Charm Semileptonic Branching Ratio}

The semileptonic branching fraction of the $c$ quark was determined with a rather large error in low energy 
experiments~\cite{clarg}. 
The measurement of this quantity at a different energy scale is of great interest due to its correlation with
the mixture of charmed hadrons in the final state; a recent OPAL measurement\cite{opal} shows excellent agreement with
the low energy results. At LEP the uncertainty on the average from the low energy
experiments $BR(c\rightarrow l)=9.8\pm0.5\%$ is an important
source of systematic error for a number of measurements based on lepton tag analyses.

\begin{figure}[htb]
\centerline{\epsfxsize 3.5truein \epsfysize 4.truein \epsfbox{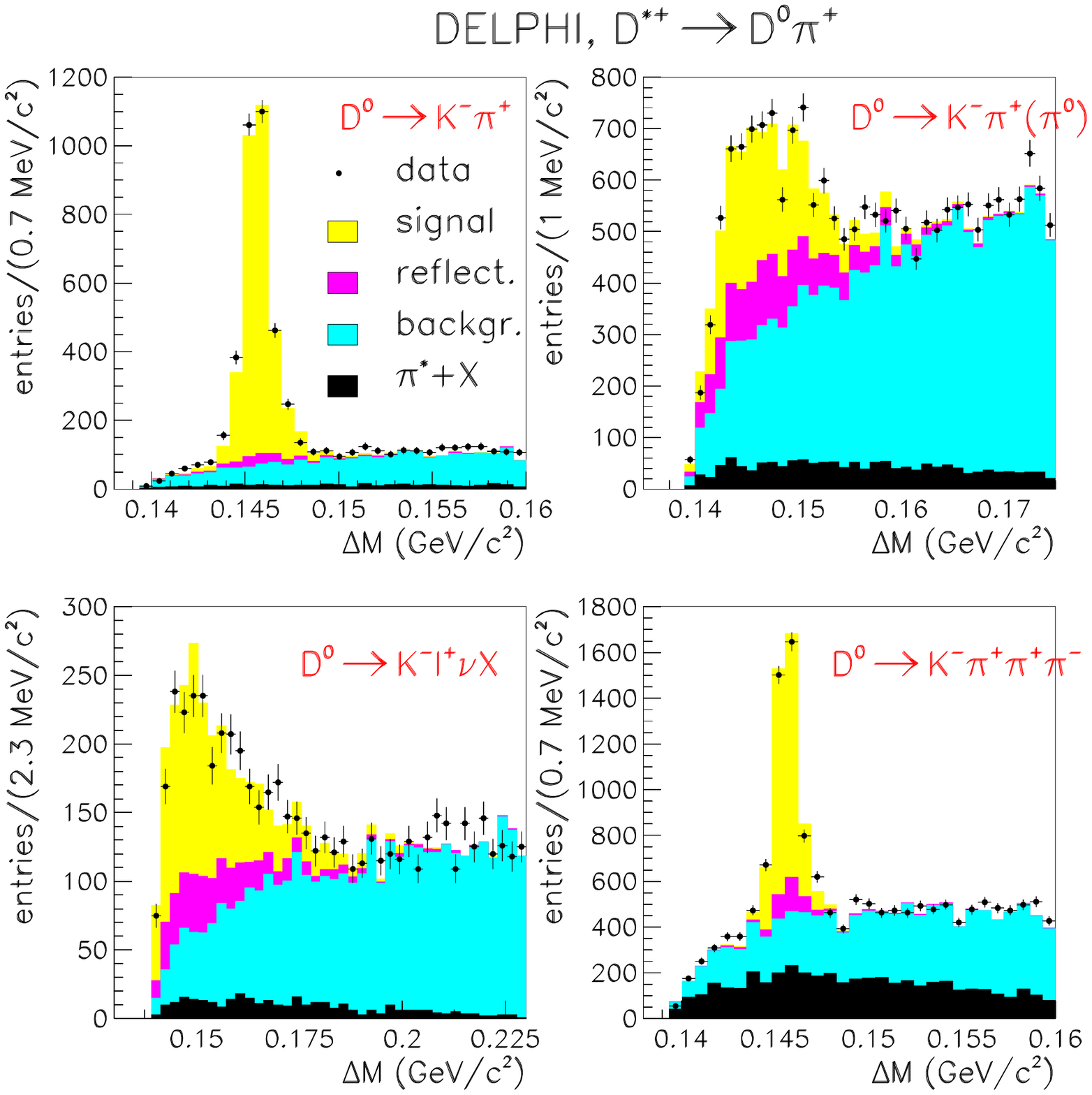}\epsfxsize 3.6truein \epsfysize 4.truein \epsfbox{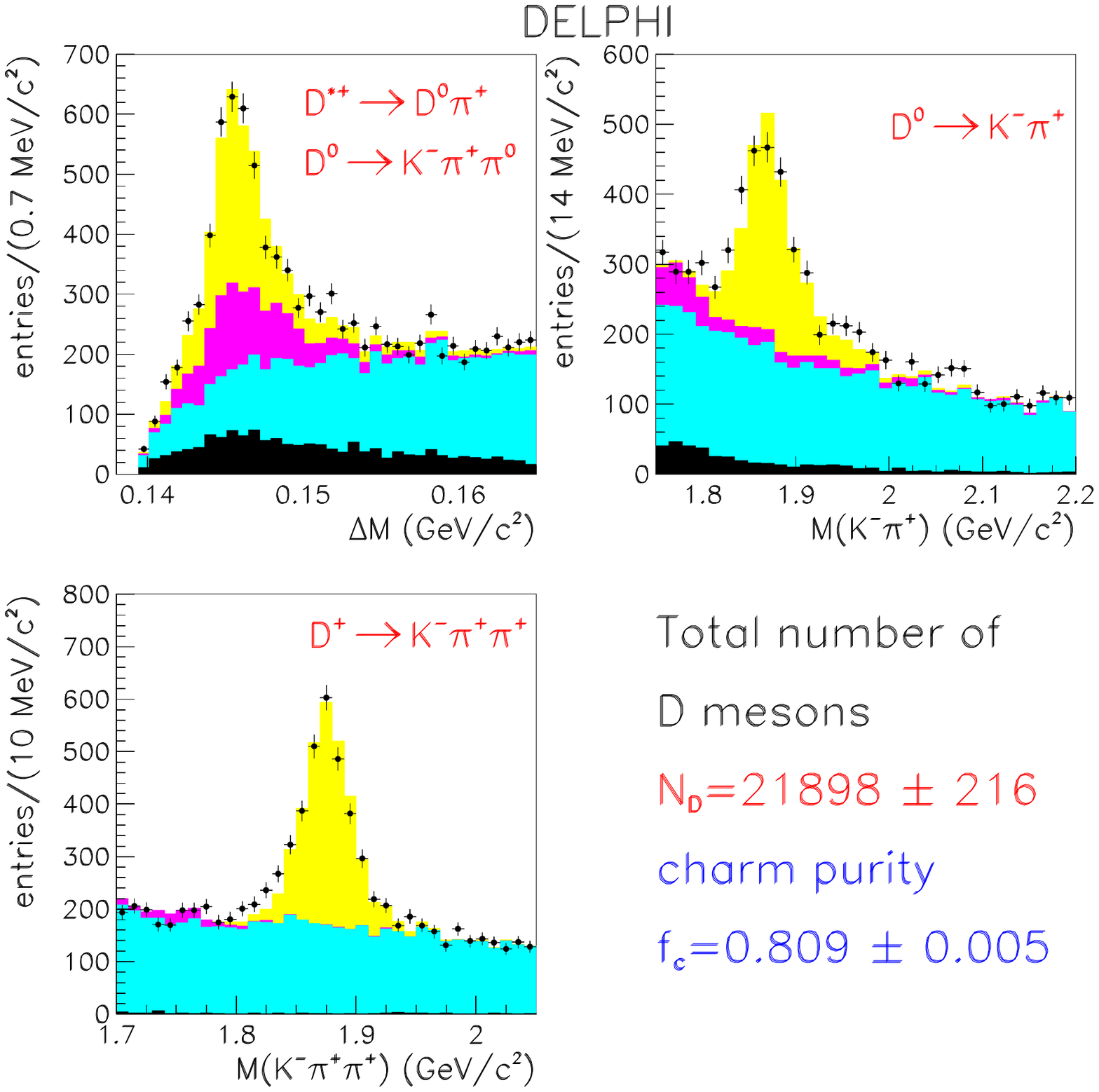}}   
\vskip -.2 cm
\caption[]{
\label{fig:d}
\small  $\Delta M$ (invariant mass) distribution for the $D^*(D^0,D^+)$ candidate decays in the different channels 
with the various contributions predicted by the simulation.}
\end{figure}

From an anti-$b$~tagged sample of hadronic $Z^0$ decays collected between 1992 and 1995, 
$c\bar{c}$ events were selected by means of an exclusive $D^*,~D^0$ and $D^+$ reconstruction in several channels. 
Charm enrichement was obtained by cutting on 
the fraction $X_E$ of the beam energy retained by the $D$ mesons. A total of $N_D=21898\pm 216$ charmed mesons were
reconstructed with a purity $f_c=80.9\pm 1.1\%$.
Figure~\ref{fig:d} shows the reconstructed $D$ meson spectra for the different channels.

The branching fraction $BR(c\rightarrow l)$ was obtained from the amount $N^l$ of the identified leptons with $P>2~GeV/c$ in 
the hemisphere opposite to the reconstructed $D$ meson and with the right charge correlation:
\begin{center}
$N^l=N^{true}_c +N^{true}_b h^{true}_b+N^{bkg}_{c} h^{bkg}_{c}+N^{bkg}_{b} h^{bkg}_{b}$.
\end{center}
In the previous expression $N^{true}_c$ is the number of the true leptons coming from $c$ decays, either directly 
produced in $Z^0\rightarrow c\bar{c}$ events or originating from the gluon splitting process $g\rightarrow c\bar{c}$ in
$Z^0\rightarrow q\bar{q}~(q=b,c)$. 
$N^{true}_b$ accounts
for the true leptons from $b$ semileptonic decays and was computed in terms of the $b$ semileptonic branching
fractions and the effective mixing which destroys the charge correlation between the lepton and the $D$ meson. 
$N^{bkg}_{c,b}$ is the sum of the contributions due to fake leptons and true
leptons from light particle decays in $c,~b$ events. 
Small corrections $h$ due to hard gluon radiation were obtained from the simulation.

\begin{figure}[htb]
\centerline{\epsfxsize 5.truein \epsfysize 2.5truein \epsfbox{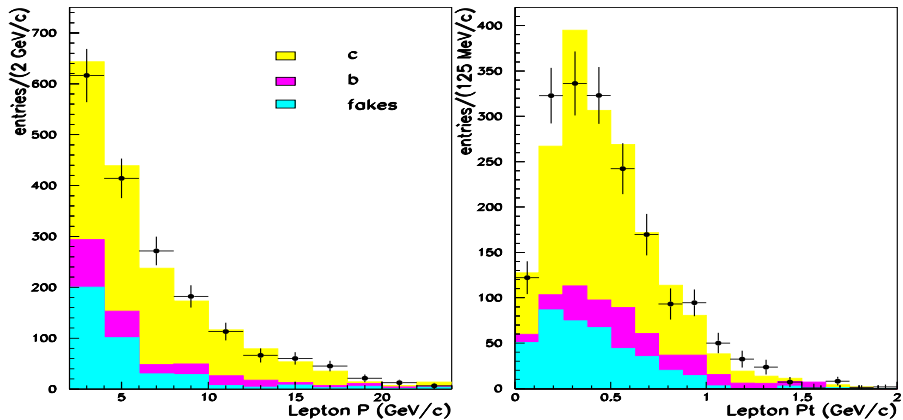}}   
\vskip -.2 cm
\caption[]{
\label{fig:lept}
\small $P$ and $P^t$ lepton candidates spectra.} 
\end{figure}

\begin{figure}[htb]
\centerline{\epsfxsize 5.truein \epsfysize 4.5truein \epsfbox{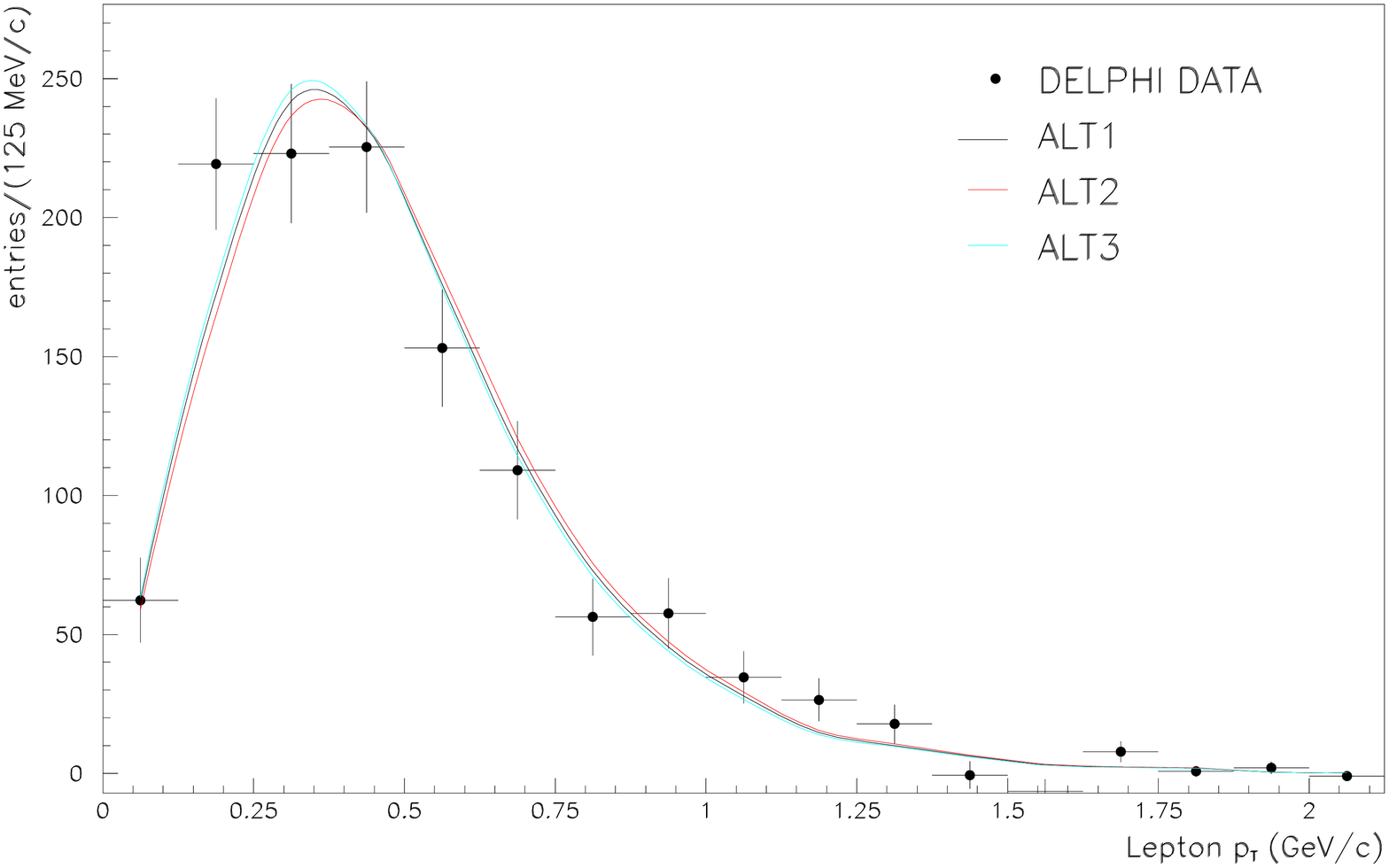}}   
\vskip -.2 cm
\caption[]{
\label{fig:spectra}
\small Reconstructed $P^t$ lepton spectrum from charm decay compared
with the predictions from the ACCM model. Labels ALT1, ALT2, ALT3 refer to different values for the strange quark mass
and the Fermi momentum according to\cite{nim}} 
\end{figure}

The lepton efficiencies were computed from the simulation
tuned to reproduce the data, taking into account the detector acceptances and found to be: $\epsilon^e=(51.4\pm1.3)\%$ and
$\epsilon^e=(60.8\pm1.3)\%$ for electrons and muons respectively. 
The number $N^{bkg}_{c,b}$ of background leptons was determined from the observed multiplicity of charged particles
in the lepton hemishepre and the misidentification probabilities $f^e_{fake}=(0.52\pm0.03)\%$ and 
$f^\mu_{fake}=(0.63\pm0.03)\%$.
The fraction of leptons from the decays of $b$ and $c$ hadrons surviving the 
momentum cut was obtained reweighting the simulation in order to reproduce the experimental values for $X_E(B)=0.702\pm0.008$ and 
$X_E(D)=0.484\pm0.008$. Different semileptonic decay models were considered according to the results reported by the LEP 
Heavy Flavour Working Group\cite{hfwg}.
After the background subtraction a yield of $N^{true}_c=1187\pm 51$ leptons from $c$ semileptonic decays 
was found. 
From the ratio $\frac{N^{true}_c}{N_D}$ the $BR(c\rightarrow l)$ was obtained to be:
\begin{center}
$BR(c\rightarrow l)=(9.58\pm0.42(stat.)\pm0.28(syst.))\%$
\end{center}
where the dominant systematics came from the lepton identification and the $c$ 
decay models. Figure~\ref{fig:lept} shows the momentum and transvers momentum spectra for the selected leptons.
Figure~\ref{fig:spectra} shows the comparison of the measured $P^t$ lepton spectrum with the 
predictions from the ACCM model with different values of the parameters used to fit DELCO and MARK III data\cite{nim}.

\section{Acknowledgments}

I am grateful to Enrico Piotto, Franco Simonetto and Jong Yi for providing me very useful 
informations.

\end{document}